\documentstyle[aps]{revtex}

\begin{document}
\title{Analysing Lyapunov spectra of chaotic dynamical systems}

\author{F.K. Diakonos\footnote{fdiakono@cc.uoa.gr}         
}
\address{
Department of Physics,
University of Athens,
GR-15771 Athens,
Greece
}

\author{D. Pingel\footnote{detlef@tc.pci.uni-heidelberg.de} and
P. Schmelcher\footnote{peter@tc.pci.uni-heidelberg.de} 
}
\address{
Theoretische Chemie,      
Institut f\"ur Physikalische Chemie,  
Im Neuenheimer Feld 229,  
69120 Heidelberg,                    
Germany               
}

\maketitle

\begin{abstract}
It is shown that the asymptotic spectra of finite-time Lyapunov
exponents of a variety of fully chaotic dynamical systems can be understood in terms
of a statistical analysis. Using random matrix theory we derive numerical and in particular
analytical results which provide insights into the overall behaviour 
of the Lyapunov exponents particularly for strange attractors.
The corresponding distributions for the unstable periodic orbits are investigated for comparison. 
\end{abstract}


Finite-time Lyapunov exponents represent an important tool for the quantitative
description of the geometrical and dynamical properties of chaotic dynamical systems \cite{Ott94}
and apply therefore to a variety of different physical situations.
Numerous recent works (see, for example, refs.\cite{Wang96,Oli96,Vulp93,Sepu89,Grass88}) try to describe
the spectrum of these exponents in terms of products of random matrices.
The majority of these investigations focus on the maximum average
Lyapunov exponent which can be estimated by taking the proper mean of an ensemble of chaotic
trajectories for a sufficiently large time interval $p$. Details of the corresponding distribution of Lyapunov exponents
(LE) are thereby not important. In order to analyze strange attractors in more detail
it is however crucial to know the spectrum of the finite-time LE which provides valuable information
on the structures and properties emerging in phase space \cite{Ott94,Kandrup97}.
For the asymptotic (large $p$) Lyapunov spectra it is known \cite{Ellis85} that,
apart from a few exceptions \cite{Prasad99}, a Gaussian approximation fits the behaviour
around the maximum very well. Very little is however known with respect to the overall behaviour of
the distribution. This is in contrast to the fact that the non-Gaussian tails of the distribution
have significant influence on physical processes \cite{Belgie93}.

In the present letter we analyze finite-time Lyapunov spectra for
low-dimensional discrete dynamical systems with fully developed chaos for the case of 
large time intervals $p \gg 1$, i.e. their asymptotic form. Our main interest is twofold. First we explore {\it common} features
of these spectra beyond their behaviour in the vicinity of their maxima and compare them
to the corresponding spectra of the unstable periodic orbits (UPOs).
It turns out that these features can be understood in terms of products of random matrices.
Second we investigate the origin of spectral properties which depend on the dynamical system.
System-dependent characteristics can e.g. be due to invariant structures in phase space like UPOs.

It is an open question how the random-like features of the chaotic dynamics determine
the distribution of the finite-time LE. One way to investigate this is to replace certain dynamical
quantities of the original system by random ones and to study the resulting changes and/or common properties in the spectra of LE.
In several recent works \cite{Genc96,Zieh99} such random-like modifications have been suggested.
Adopting this methodology here, we will use four different ensembles of trajectories leading
to different distributions for the corresponding LE. Each ensemble $E_i$ consists of
$N$ trajectories (typically $N \approx 10^6$) of a given length $p$ ($p \gg 1$)
and is given by (i) the distribution of the starting points of the finite trajectories and
by (ii) the rule applied to generate the trajectory itself. We will then be primarily 
interested in the distributions of the finite-time maximum LE $\lambda = \frac{1}{p} \ln \Lambda$
belonging to these ensembles $\{E_i\}$. $\Lambda$ is the largest of the absolute values of the eigenvalues belonging to 
the transfer matrix $M^{(p)}_i=\displaystyle{\prod_{k=1}^p} M_i(\vec{x}_{k})$ with 
$\{\vec{x}_{k}|k=1,...,p\}$ being a finite-time trajectory of the ensemble $E_i$.
$M_i(\vec{x}_{k})$ is the stability matrix belonging to the (chaotic) dynamical law
$\vec{x}_{n+1}=\vec{F}(\vec{x}_n)$.  The index $i$ of $M_i$ indicates the ensemble $E_i$ according to which 
the points $\{\vec{x}_{k}\}$ of the trajectories are determined. It is important to note
that all ensembles $\{E_i\}$ (see below) use the specific functional form of the stability matrix
belonging to $\vec{F}$ but involve trajectories with different degrees of randomness.
In the following we specify the ensembles $\{E_i\}$. 

$E_1$ consists of trajectories of length $p$ obtained via iteration of the chaotic dynamical 
law $\vec{F}$. The corresponding initial conditions are distributed according
to the invariant density of the map. This gives the so-called finite-time Lyapunov exponent distribution (FTLED)
(see ref.\cite{Ott94} and refs. therein). The ensemble $E_2$ consists of trajectories which are generated by a random variable
distributed according to the invariant density of the chaotic map $\vec{F}$.
This yields the bootstrap Lyapunov exponent distribution (BLED) \cite{Genc96,Zieh99}.
Compared to the FTLED the BLED corresponds to a dynamics with enhanced random character.
The successive points of the bootstrap trajectory are completely uncorrelated.
The third ensemble $E_3$ uses a uniformly distributed random variable for the
generation of the trajectories. The range of the uniform distribution is chosen according
to the phase space of the dynamical system. This case corresponds to a random
matrix simulation of the dynamical system however respecting the form of the stability matrix ${\bf{M}}$ 
which belongs to the map $\vec{F}$. The resulting distribution of the Lyapunov exponents
is called the random matrix Lyapunov exponent distribution (RMLED). Within the
present investigation the ensemble $E_3$ possesses the highest degree of randomness.
The fourth ensemble $E_4$ consists of the UPOs of the dynamical system $\vec{F}$ with period $p$
and correspondingly the distribution of their maximal LE. For fixed $p$ the
number of trajectories contained in $E_4$ is finite according to the
topological entropy of the corresponding phase space.

Let us now explore the distributions defined above for a variety
of low-dimensional fully chaotic systems. Our main goal is to analyse and understand the
overall behaviour of the FTLED for these systems. 
We begin with a simple $1d$ example: the logistic map.
Results on the FTLED for this system can be found in ref.\cite{Prasad99}.
Noteworthy the FTLED has a non-Gaussian form with one dominating central cusp.
In comparison to this our numerical calculations on both the BLED and RMLED show that
they are smooth functions with a Gaussian-like maximum but with characteristic asymmetric tails.
For the maximum Lyapunov exponent distribution of the UPOs the exact result is
a $\delta$-function i.e. $\rho_p(\lambda)=\delta(\lambda - \ln 2)$ independent
of the period $p$. Looking at the FTLED one observes that the tails of the
distribution are rather similar to the tails of the BLED and the RMLED while
the cusp (maximum) is located exactly at $\lambda=\ln 2$.
The BLED (or RMLED) reflects therefore the overall behaviour of the
FTLED i.e. describes the envelope of the FTLED. The latter possesses an
additional central peak at the position of the Lyapunov exponents of the UPO's.
These features will in the following turn out to be common for a broad class of
dynamical systems.

We focus in the following on two-dimensional systems with a strange attractor.
The Henon map \cite{Henon76} for $a=1.4$ and $b=0.3$ and
the Ikeda map \cite{Ikeda85} for $a=0.9$ and $b=6$ are prototypes of such systems.
The FTLED of both are presented in Fig.~1. The ensemble $E_1$
consists of $10^6$ trajectories of length $p=27$ for both the Henon
and Ikeda map. The results are surprisingly similar to the $1d$ case: A characteristic
envelope with asymmetric tails dominates the distribution
while superimposed peaks indicate the presence of additional structures in phase space.
As we shall see in the following three regions with different functional forms of the envelope can be distinguished
in the FTLED as well as the RMLED and BLED:
A fast asymptotic algebraic decay for large values of $\lambda$,
a Gaussian-like behaviour around the maximum and a dominating exponential decay for sufficiently small $\lambda$.
Figure 2 shows the BLED and RMLED for these maps for the same length $p$.
The position of the maximum of the RMLED is sensitive with respect to the random number intervals
chosen, i.e. its location carries the information of the position of the attractor in phase space.
Based on the above results and observations we are naturally lead to the following conclusion:
the basic possible features of the smooth envelope of the FTLED (asymmetric structure, asymptotic tail properties)
of a chaotic dynamical system are of random origin and can be obtained and understood by a corresponding study of
random matrices (see below). Additional superimposed structures are signatures of e.g. invariant sets in 
phase space and are therefore of exclusively deterministic dynamical origin.

To elucidate and quantify the above observations we perform in the following an analytical investigation
of the RMLED. This will allow us to thoroughly understand the behaviour of the RMLED and consequently the
corresponding aspects of the FTLED. We begin by introducing a fictitious dynamical system with a stability
matrix ${\bf{M}}_i$ of strongly random character i.e. 
 $$\bbox{M}_i= r_i \bbox{A} =
\left( \begin{array}{cc} {\rm{a}} r_i & {\rm{b}} r_i \\
{\rm{c}} r_i & {\rm{d}} r_i \end{array} \right) $$
where $r_i$ is a random variable uniformly distributed in $[0,R]$ and $i$ labels the fictitious trajectory
of length $p$. The simple form of the matrix $\bbox{M}_i$ allows us to factorize the random
variables $\{r_i| i=1,...,p\}$ of the transfer matrix $\bbox{M}^{(p)}=\displaystyle{\prod_{k=1}^p \bbox{M}_k}$
and to reduce the problem of the product of random matrices to that of a product of random numbers.
The matrix structure is then retained in the constant matrix ${\bf{A}}$ which is assumed to be nonsingular.

The distribution of the maximum LE for trajectories of length $p$ of this system is determined as:
\begin{equation}
\rho_p(\lambda)=\int_0^R dr_1~\int_0^R dr_2~...~\int_0^R dr_p ~
\delta(\lambda-\frac{1}{p} \ln \displaystyle{\prod_{i=1}^p} r_i \vert
\Lambda_{max} \vert)~\displaystyle{\prod_{l=1}^p} \tilde{\rho}(r_l)
\label{eq:toy1}
\end{equation}
where $\tilde{\rho}(z)=\Theta(z) \Theta(R-z) \frac{1}{R}$, $\Theta$ being the
step function and $\Lambda_{max}$ is given by
$\Lambda_{max}=\frac{1}{2} ({\rm{Tr}}\bbox{A} +
{\rm{sign}}({\rm{Tr}}\bbox{A})\sqrt{({\rm{Tr}}\bbox{A})^2 - 4 {\rm{det}}\bbox{A}})$.
Using the substitution $t_i=\ln (r_i \vert \Lambda_{max} \vert)$ and
performing the Fourier transform of the $\delta$-function involved in
(\ref{eq:toy1}) we obtain:
\begin{equation}
\rho_p(\lambda)=\frac{p}{2 \pi (R \vert \Lambda_{max} \vert)^p}
\int_{-\infty}^{\infty}
dk~e^{-i k p \lambda} \int_{-\infty}^{\ln(R \vert \Lambda_{max} \vert)} dt_1
....\int_{-\infty}^{\ln(R \vert \Lambda_{max} \vert)} dt_p~
exp\left({(1+ i k) {\displaystyle{\sum_{i=1}^p}} t_i}\right)
\label{eq:toy2}
\end{equation}
The integrations over $t_i$ in (\ref{eq:toy2}) can be easily performed
leading to a $p$-th order pole in the complex $k$-plane at $k=i$. This
pole structure, corresponding to values of $k$ for which the exponent
of the second exponential term in (\ref{eq:toy2}) vanishes and for which the integration of $t_i$ leads
to singularities, is responsible for the features of the LED described above.
Complex integration finally yields:
\begin{equation}
\rho_p(\lambda)=\frac{p^p}{(p-1)!} \left( \ln(R \vert \Lambda_{max} \vert)
- \lambda \right)^{p-1} e^{-p(\ln(R \vert \Lambda_{max} \vert) - \lambda)}~
\Theta(\ln(R \vert \Lambda_{max} \vert) - \lambda)
\label{eq:toy3}
\end{equation}
Eq.(\ref{eq:toy3}) demonstrates that the exponential behaviour
dominates for sufficiently small values of $\lambda$. Around the maximum 
at $\lambda_o=\ln(R \vert \Lambda_{max} \vert)-(1-\frac{1}{p})$
the saddle point approximation ($p \gg 1$) gives
a Gaussian. For values of $\lambda$ close to the maximum value
$\lambda_{max}=\ln(R \vert \Lambda_{max} \vert)$ we arrive at a power law
behaviour, i.e. an algebraic decay. 

Although the fictitious dynamical model discussed above captures the main features
of the statistical properties of the distributions of Lyapunov exponents
it is clearly desirable to investigate chaotic dynamical systems
for which the RMLED can be obtained analytically. To this end
let us consider the dynamical system defined by the quadratic equations:
$x_{n+1}=a y_n^2 + b~;~y_n=c x_n + d$. 
This system possesses for $a=c=d=1$ and $b=-2.5$ a strange attractor
which contains repeating crosses of decreasing size.
For that reason we call it in the following the cross map.
The average maximum Lyapunov exponent of this attractor is $\bar{\lambda} \approx 0.123$
while its fractal dimension is $d_F \approx 1.78$.
The RMLED for the cross map can be calculated analytically following the line described
above for our fictitious model. One pecularity of the cross map is that
one has to distinguish between the RMLED obtained through trajectories
with even and odd length $p$. The reason is that for odd $p$
both eigenvalues of the transfer matrix $M^{(p)}$ have the same absolute
value, while for even $p$ there are two eigenvalues different with respect to their absolute value 
and one has to select the maximal one. After a tedious calculation
using complex contour integration techniques we find for the RMLED of the cross map 
for odd values of $p$ the result:
\begin{equation}
\rho_p(\lambda)=\frac{2p}{p!} \left[ \frac{p R_2}{R_2-R_1}
\right]^{p}~exp\left({-p(\Delta - 2\lambda)}\right)~
{\displaystyle{\sum_{j=0}^{p}}}  \left(
\begin{array}{c} p \\ j \end{array} \right) (\Delta-\frac{j}{p} \ln\vert
\frac{R_2}{R_1} \vert - 2\lambda)^{p - 1} 
~\Theta(\Delta-\frac{j}{p} \ln\vert \frac{R_2}{R_1} \vert - 2\lambda)
\label{eq:odd}
\end{equation}
Here we have taken the random variables appearing in the stability matrix of the cross map to be uniformly distributed
in the interval $[R_1,R_2]$. The parameter $\Delta$ is given as:
$\Delta= \ln\vert 2 a c R_2 \vert$. From eq.(\ref{eq:odd}) we see that the RMLED of the cross map
is essentially a product of a single exponential and a sum of
power laws. The latter possess all the same power $p-1$ and differ
only with respect to the constants involved. The similarity to the RMLED
result of our model system in eq.(\ref{eq:toy3}) is obvious which
confirms the universality of certain features of the RMLED.
The RMLED for the case of even $p$ is given by the integral:
\begin{equation}
{\rho}_p(\lambda)= 2 \rho_{\frac{p}{2}}(\lambda) \int_{-\infty}^{\lambda} dz~
\rho_{\frac{p}{2}}(z)
\label{eq:even}
\end{equation}
with $\rho_{\frac{p}{2}}(x)$ according to eq.(\ref{eq:odd}). The integration in
(\ref{eq:even}) can be performed analytically leading to a lengthy expression
which will be not given here. The main characteristics of the function
${\rho_p}(\lambda)$ are again the features stated previously:
a dominating exponential behaviour for sufficiently small values of $\lambda$, a Gaussian maximum
and a fast algebraic decay for $\lambda$ close to its maximum value. 
We have also studied the FTLED, BLED and the distribution of the Lyapunov exponents of
the UPOs for the cross map for $p=28$. The results are shown in Fig.~3.
The envelope of these distributions exhibit the same features as discussed above.
The additional structures present in the FTLED of the cross map are, as can be 
seen from Fig.~3, due to the presence of the UPOs which
provide signatures of deterministic dynamical origin.

Finally let us consider the distributions of the Lyapunov exponents of the UPO's of the Henon ($p=27$) and the
Ikeda ($p=14$) map \cite{Schmel97,Diak98} which are presented in Fig.~4. 
The signs of the characteristic properties of the envelopes discussed above are also 
visible here. An interesting feature appears for the distribution of the LE of the UPOs of the Ikeda map:
it is shifted significantly compared to the FTLED.
This shift is probably due to the fact that the UPOs fail to reproduce the invariant
density of the attractor, at least up to the above-considered period.
It is well-known that the Ikeda attractor needs a description going beyond the linear neighbourhood \cite{Abarba91}.

Summarizing our results we have demonstrated that the overall behaviour of the finite-time Lyapunov exponent distributions of fully
chaotic dynamical systems show general characteristics i.e. they can be understood in
terms of statistical random matrix simulations of the systems. Seemingly this holds also
for the distributions of the Lyapunov exponents of the unstable periodic orbits embedded into the
chaotic phase space. Since Lyapunov spectra are at the heart of our understanding of chaotic systems
in general our results apply to a variety of different physical systems.

Financial support by the Deutsche Forschungsgemeinschaft and the Landesgraduiertenf\"orderungsgesetz
Baden-W\"urttemberg (D.P.) are gratefully acknowledged. We appreciate valuable discussions with O.~Biham.

\vspace*{1.0cm}

\begin{center}
{\large{\bf{Figure Captions}}}
\end{center}
\vspace*{-0.5cm}

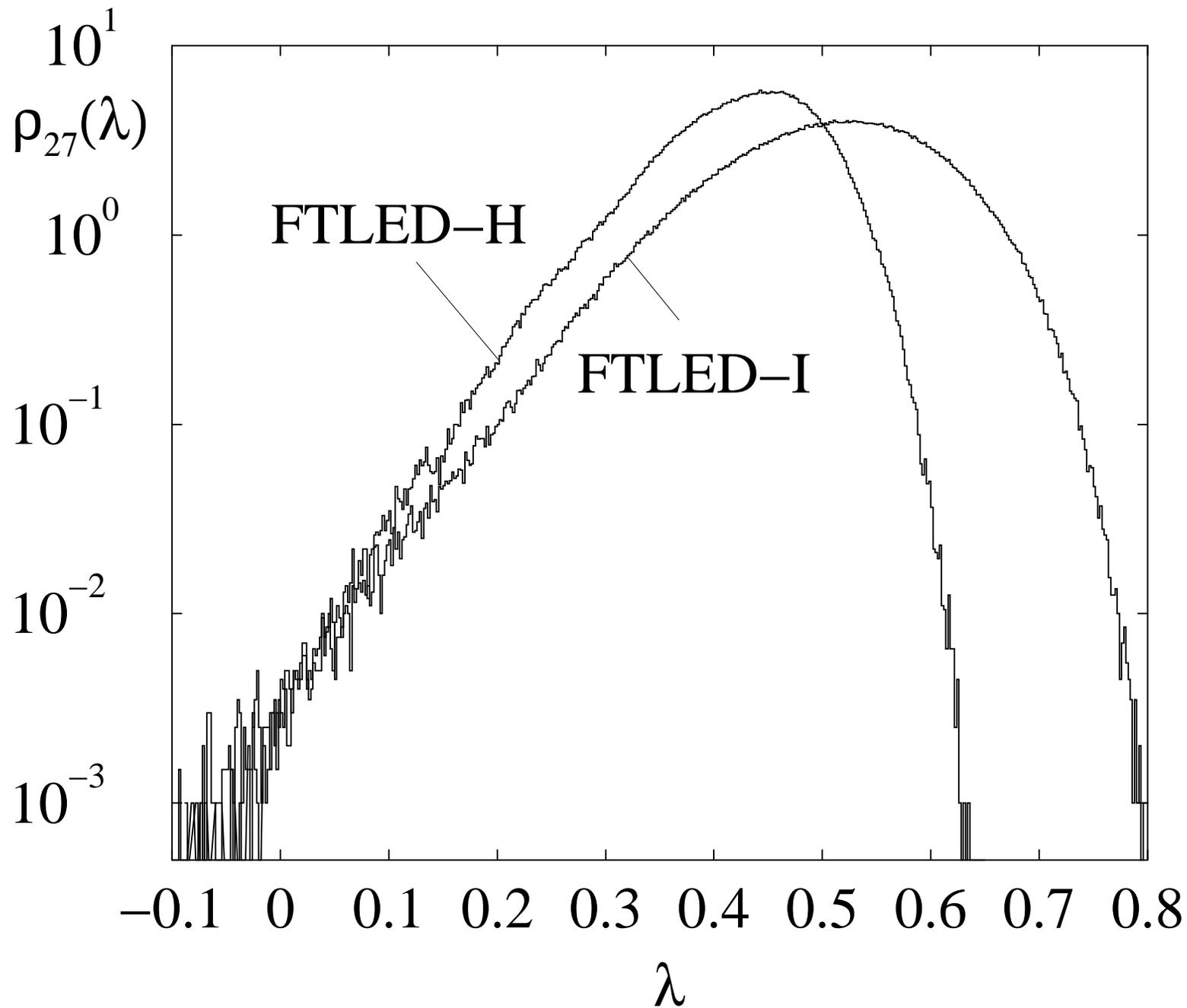
\begin{figure}
\caption{The FTLED (see text) of the Henon and Ikeda map for $p=27$.}
\end{figure}
\vspace*{-0.5cm}

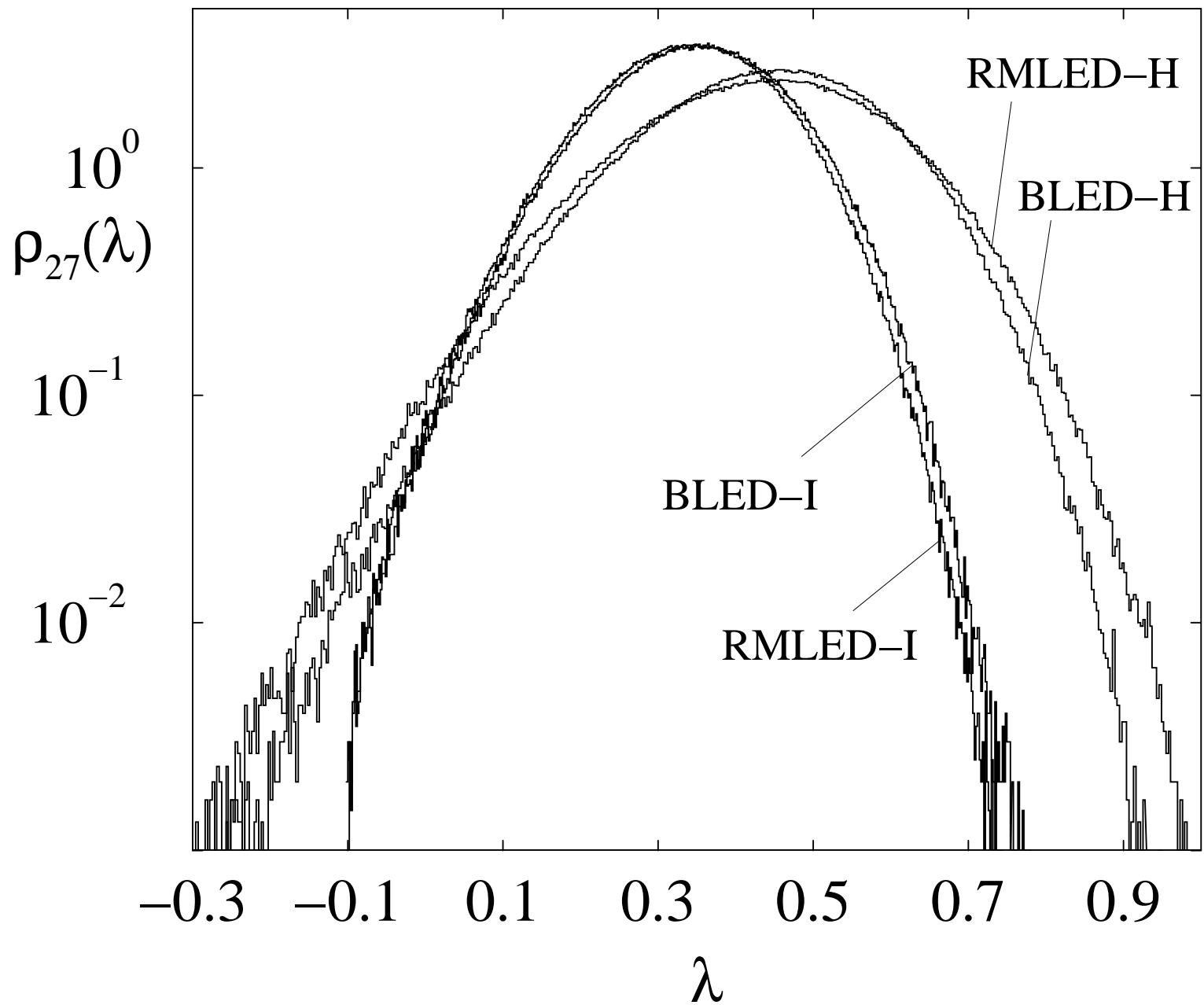
\begin{figure}
\caption{The BLED and RMLED (see text) of the Henon (-H) and Ikeda (-I) map for $p=27$.}
\end{figure}
\vspace*{-0.5cm}

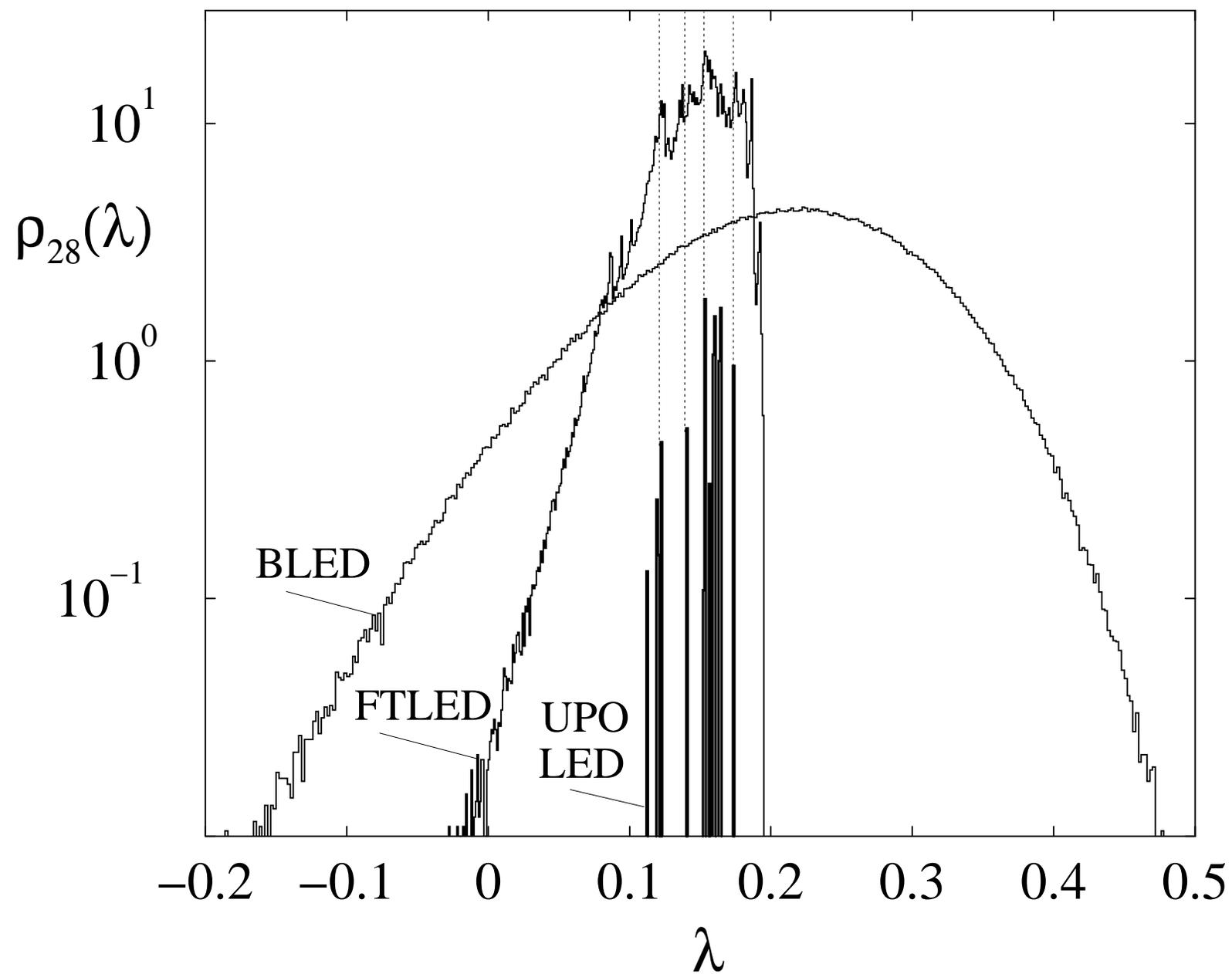
\begin{figure}
\caption{The FTLED, BLED and the distribution of the Lyapunov exponents of the unstable
periodic orbits for the cross map for $p=28$.}
\end{figure}
\vspace*{-0.5cm}

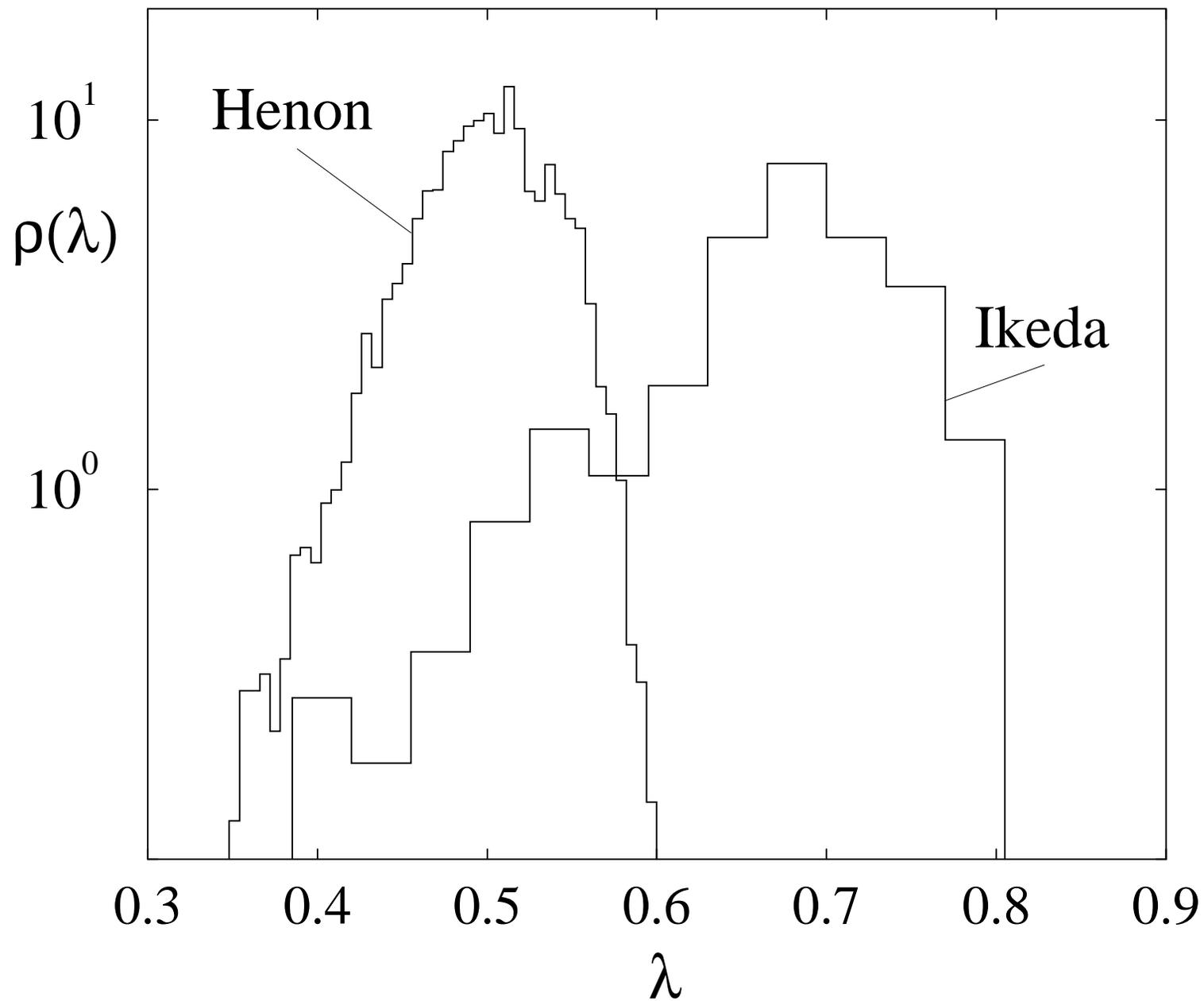
\begin{figure}
\caption{The distribution of the Lyapunov exponents of the unstable periodic orbits of the Henon ($p=27$) and Ikeda map ($p=14$).}
\end{figure}
\vspace*{-0.5cm}

\end{document}